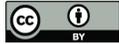

Earth System Dynamics

# Seasonality of the hydrological cycle in major South and Southeast Asian river basins as simulated by PCMDI/CMIP3 experiments

S. Hasson[1,2], V. Lucarini[1,3,4], S. Pascale[1], and J. Böhner[2]

[1]KlimaCampus, Meteorological Institute, University of Hamburg, Hamburg, Germany
[2]KlimaCampus, Institute of Geography, University of Hamburg, Hamburg, Germany
[3]Department of Mathematics and Statistics, University of Reading, Reading, UK
[4]Walker Institute for Climate System Research, University of Reading, Reading, UK

*Correspondence to:* S. Hasson (shabeh.hasson@zmaw.de)

**Abstract.** In this study, we investigate how PCMDI/CMIP3 general circulation models (GCMs) represent the seasonal properties of the hydrological cycle in four major South and Southeast Asian river basins (Indus, Ganges, Brahmaputra and Mekong). First, we examine the skill of the GCMs by analysing their performance in simulating the 20th century climate (1961–2000 period) using historical forcing (20c3m experiment), and then we analyse the projected changes for the corresponding 21st and 22nd century climates under the SRESA1B scenario. The CMIP3 GCMs show a varying degree of skill in simulating the basic characteristics of the monsoonal precipitation regimes of the Ganges, Brahmaputra and Mekong basins, while the representation of the hydrological cycle over the Indus Basin is poor in most cases, with a few GCMs not capturing the monsoonal signal at all. While the model outputs feature a remarkable spread for the monsoonal precipitation, a satisfactory representation of the western mid-latitude precipitation regime is instead observed. Similarly, most of the models exhibit a satisfactory agreement for the basin-integrated runoff in winter and spring, while their spread is large for the runoff during the monsoon season. For the future climate scenarios, most models foresee a decrease in the winter $P - E$ over all four basins, while agreement is found on the decrease of the spring $P - E$ over the Indus and Ganges basins only. Such decreases in $P - E$ are mainly due to the decrease in precipitation associated with the western mid-latitude disturbances. Consequently, for the Indus and Ganges basins, the runoff drops during the spring season while it rises during the winter season. Such changes indicate a shift from rather glacial and nival to more pluvial runoff regimes, particularly for the Indus Basin. Furthermore, the rise in the projected runoff, along with the increase in precipitation during summer and autumn, indicates an intensification of the summer monsoon regime for all study basins.

## 1 Introduction

Substantial anthropogenic climate-change-driven changes in the global hydrological cycle (Held and Soden, 2006; Allan, 2011) will largely impact the spatial and temporal pattern of water supply on a regional and global scale. Since almost any human activity, and in particular agriculture and industry, strongly depend on water availability, additional pressures on the ongoing economic development and population growth will be associated with such changes (Kundzewicz et al., 2008) and may particularly be strong in areas more vulnerable to drought or flood. The situation is expected to be especially critical for highly populated regions of South and Southeast Asia, whose agriculture-based economies and rapidly developing industrial systems are largely dependent on variable water supplies. Therefore, inferring detailed information about the climate change, its impact on the water resources, and its consequent implications for the socio-economic development sectors is vital for adequate adaptation and mitigation policies in the region.

Despite their structural limitations and ambiguities in the values of crucial parameters (Held and Soden, 2006; Lucarini et al., 2008), general circulation models (GCMs) are presently the most powerful tools for simulating the Earth's climate, its natural variability and the impact of anthropogenic forcing. The GCM simulations under diverse scenarios are used by various scientific communities to inform stakeholders and policymakers on key impacts of climate change and to support the development of efficient mitigation and adaptation policies (IPCC AR4, 2007). In particular, GCMs are extensively being used to understand the effects of global warming on the water cycle at a global and a regional scale. It is widely accepted that a realistic representation of the hydrological cycle is, however, non-trivial in these models because the hydro-meteorological processes take place on a vast range of time- and space scales, including regimes that can be represented only through parameterizations (Hagemann et al., 2006; Tebaldi and Knutti, 2007). Biases due to processes occurring at very small scales can have global impacts: Liepert and Previdi (2012) and Liepert and Lo (2013) have shown that most of the Climate Model Inter-Comparison Project Phase-3 (CMIP3 – Meehl et al., 2007) and Phase-5 (CMIP5 – Taylor et al., 2012) GCMs have serious problems in conserving the global water mass and that such inconsistencies are truly macroscopic for few models. Such an inconsistent representation of the hydrological cycle causes further biases in the energetics of the climate models (Lucarini and Ragone, 2011; Liepert and Previdi, 2012; Liepert and Lo, 2013), and leads to significant uncertainties in the climate-change-induced variations of the global and regional hydrological regimes.

The hydro-climatology of South and Southeast Asia is extremely complex, as the precipitation regimes associated with the large-scale circulations (monsoonal systems and mid-latitude disturbances), the local scale moisture availability deriving from intense evapotranspiration due to irrigation (Saeed et al., 2009), and meltwater contribution from the existing cryosphere are involved in determining the seasonality of the overall hydrological cycle, with vast differences across the region. Since climate change involves variations in the geographical distribution, timing and intensity of the South Asian and Southeast Asian summer monsoons and the extratropical cyclones, it is crucial to realistically represent such large-scale weather systems in order to have a satisfactory representation of the hydrological cycle by the climate models and to be able to have a good representation of the effects of climate change on the hydrology of the region. The studies performed in such regards have all demonstrated the model biases in the spatio-temporal distribution and magnitude of the summer monsoonal precipitation (Goswami, 1998; Lal et al., 2001; Kang et al., 2002; Annamalai et al., 2007; Lin et al., 2008, Boos and Hurley, 2013). In a recent study, Sperber et al. (2013), by comparing the CMIP3 and CMIP5 model outputs, show that no single model represents all the relevant aspects of the monsoon. They also found that the models see serious difficulties in simulating the timings of the monsoon onset as compared to the timings of the peak and the withdrawal.

As a result, considerable uncertainties exist on the projected changes for the hydrology of South and Southeast Asia. For instance, Arnell (1999), by using the Hadley Centre climate models (HADCM3), suggested an increase in the annual runoff for the Asian and Southeast Asian region, whereas Arora and Boer (2001), by using the Canadian Centre for Climate Modelling and Analysis (CCCma) coupled climate model, suggested instead a decrease in it for the Southeast Asian river basin. Linking such contrasting findings with the individual model biases, Nohara et al. (2006) assessed the projected changes in the hydrology of 24 major rivers using the weighted ensemble mean approach from 19 climate models, indicating difficulties of reproducing the observed discharges by any single model.

As a side note, we wish to remark that multi-model ensemble mean climatological estimators should be used with caution, as these are in general, ill-defined statistical quantities. The outputs of different models do not form a statistically homogeneous object, given by equivalent realizations of a process. In general, there is no reason to expect that the biases of different models tend to cancel out in a fashion similar to what once may expect from assuming simplistically a sort of law of large numbers. Considering the mean as approximating the "truth" and the standard deviation as describing the uncertainty could be misleading (Lucarini et al., 2008; Hasson et al., 2013a). For any given diagnostic target variable, the multi-model ensemble estimates do not necessarily outperform any single best model (Tebaldi and Knutti, 2007) and largely depend upon the skill of the ensemble members featuring huge structural and physical differences. Sperber et al. (2013) have shown that the CMIP3 multi-model ensemble mean outperforms each single model only for three diagnostics – precipitation, 850 hPa wind and peak month – out of the seven considered in their study. It has been noted from their calculated metrics that the multi-model ensemble does not give a better indication than the best performing model for a given metric when the models feature a weak inter-model agreement (e.g. the onset). Computing the multi-model averages for climate change signals raises further conceptual problems, because qualitatively different responses from different models are averaged out, and information on the dynamical processes contributing to determining the climate response is lost, so that the quality check of the results becomes extremely difficult. Furthermore, ensemble-averaged results are of little relevance if one wants to provide information on the best practices for downscaling, since in this case a specific GCM or a few of them can be considered.

Boos and Hurley (2013), attributed the bias in the thermodynamic structure of the summer monsoon as represented by the CMIP3 and CMIP5 climate models to an inaccurate representation of the Karakoram–Himalayan orography, which results in negative precipitation anomalies over the Indian region. These findings are in agreement with recent results of the authors, who recently investigated the representation of mean annual hydrological cycles of the four major river basins of South and Southeast Asia (Indus, Ganges, Brahmaputra and Mekong) by the CMIP3 GCMs, reporting specific information for each of the analysed GCM (Hasson et al., 2013a). The results suggest the presence of a systematic underestimation of $P - E$ for all basins, mainly due to the underestimation of precipitation by the 20th century. Additionally, looking at climate projections under the SRESA1B scenario, the analysis of the GCM results suggests an increase (decrease) in $P - E$ for the Ganges, Brahmaputra and Mekong basins (Indus Basin) and increase (decrease or no change) in the risk of hydro-meteorological extremes for the Ganges and Mekong (Indus and Brahmaputra) basins by the 21st and 22nd centuries. These findings match well with GCM predictions about the impact of greenhouse gases forcing on the main weather systems of the region, which foresee an intensification of the South Asian summer monsoonal precipitation regime, together with an early onset and a weakening of the monsoon circulation winds (Turner and Annamalai, 2012); the Mediterranean storm track, however, is expected to weaken and shift polewards (Bengtsson et al., 2006), thus reducing its moisture input to the region.

As the major weather systems influencing the region (summer monsoon and extratropical cyclones) feature a clear seasonal cycle of their associated precipitation regimes, analysing the annual cycles of the simulated quantities seems necessary in order to assess how well the basic features of these regimes are reproduced by the models. With this goal in mind, the present study investigates the intra-annual distribution of precipitation, evaporation and runoff for the South and Southeast Asian river basins (Indus, Ganges, Brahmaputra and Mekong), using the output from the same set of climate models considered in Hasson et al. (2013a), for the 1961–2000 period and for the climate scenario SRESA1B. Hasson et al. (2013a) show that a few CMIP3 AOGCMs feature serious water balance inconsistencies for 20th century climate (1961–2000) over some of the considered river basins. Here, presenting the intra-annual investigations of the hydrological cycle, we discuss a possible link between such model inconsistencies and the misrepresentation of the seasonal water cycle. We also test whether the models featuring realistic annual averages of the main hydrological basin-integrated quantities feature inconsistencies on the intra-annual scale. The projected seasonal changes in the hydrological cycle under SRESA1B scenario for the 21st and 22nd centuries, considered in the second part of this study, provide additional indications with respect to what is presented in Hasson et al. (2013a) about the possible future scenarios for the hydrology of the region. In a following study, we will extend our investigations to a recently available data set of the CMIP5 climate models in order to assess how these models represent the regional hydrological cycle after going through an extensive development, introducing higher resolutions, atmosphere and land use and vegetation interaction, detailed aerosols treatment, carbon cycle, etc. (Taylor et al., 2012). Our present findings will serve as a benchmark, providing an opportunity to see how the newly introduced features and enhanced processes, now implemented in several CMIP5 climate models, have impacted the representation of the hydrological cycle over the region.

The paper is structured as follows. In Sect. 2 we discuss the river basins considered in this study and the basic characteristics of their hydrology. In Sect. 3, we briefly describe the CMIP3 simulations used in the analysis and the methodology adopted in order to compute the hydrological quantities. In Sect. 4, we present the performance of the CMIP3 coupled climate model simulations in terms of their skill in reproducing the intra-seasonal variations for the historical climate, and in Sect. 5, we present the seasonal changes in the same hydrological quantities for the future climates of the 21st (2061–2100) and 22nd (2161–2200) centuries under SRESA1B scenario (720 ppm of $CO_2$ after 2100). Section 6 summarizes the main results of this study and presents perspectives for future investigations.

## 2 Study area

The study area includes four major river basins of South and Southeast Asia, namely the Indus, Ganges, Brahmaputra and Mekong. These basins are roughly included between 5°–40° N and 60°–110° E (Fig. 1). Their hydro-climatology is mainly determined by two different large-scale climatic features, the South and Southeast Asian summer monsoon and the western (predominantly winter) mid-latitude disturbances, and their interactions with the local and sub-regional forcing.

The Asian monsoon system has three different, but interrelated components: South Asian, Southeast Asian and East Asian monsoons (Janowiak and Xie, 2003). Three of our study basins (Indus, Ganges and Brahmaputra basins) are mainly influenced by the South Asian monsoon, whereas the Mekong Basin comes under the influence of the Southeast Asian monsoon. The Asian monsoon is generally a thermally driven, large-scale weather system associated with a temperature gradient between land and ocean (Clift and Plumb, 2008) and with the formation of a warm anticyclone in the middle-upper troposphere ("Monsoon High"), centred above the upper Tsangpo depression (Böhner, 2006). Generally with an instant onset, the monsoon precipitation starts over the Mekong Basin in mid-May, then it extends towards the northwest, reaching the Brahmaputra, Ganges and Indus basins by mid-June to July (Fasullo and Webster, 2003), and features a fairly smooth retreat in October (Goswami, 1998). The onset of the monsoon is characterized by an abrupt increase in the daily rainfall, e.g. from below 5 mm day$^{-1}$ to over 15 mm day$^{-1}$ over India, which persists throughout the monsoon season, whereas the retreat of the monsoon denotes a reversal to the dry, dormant conditions (Fasullo and Webster, 2003). Intense precipitation continues throughout the summer monsoon season, although often interrupted by sudden breaks (Ramaswamy, 1962; Miehe, 1990; Böhner, 2006). The onset of the monsoon and the duration of the summer breaks are both critical factors, especially for the area of rain-fed agriculture, where crops are extremely sensitive to delays in the start of the rainy season or to prolonged dry periods during the summer.

**Table 1.** List of GCMs used in the study. These constitute the subset of all GCMs included in the PCMID/CMIP3 project providing all the climate variables of our interest.

| Name (Reference) | Institution | Grid Resolution (Lat × Lon) |
| --- | --- | --- |
| CNRM-CM3 (Salas-Mélia et al., 2005) | Météo-France/Centre National de Recherches Météorologiques, France | T63 |
| CGCM2.3.2 (Yukimoto and Noda, 2002) | Meteorological Research Institute, Japan Meteorological Agency, Japan | T42 |
| CSIRO3 (Gordon et al., 2002) | CSIRO Atmospheric Research, Australia | T63 |
| ECHAM5 (Jungclaus et al., 2006) | Max Planck Institute for Meteorology, Germany | T63 |
| ECHO-G (Min et al., 2005) | MIUB, METRI, and M&D, Germany/Korea | T30 |
| GFDL2 (Delworth et al., 2005) | US Dept. of Commerce/NOAA/Geophysical Fluid Dynamics Laboratory, USA | 2.5° × 2.0° |
| GISS-AOM (Lucarini and Russell, 2002) | NASA/Goddard Institute for Space Studies, USA | 4° × 3° |
| INMCM3 (Volodin and Diansky, 2004) | Institute for Numerical Mathematics, Russia | 5° × 4° |
| IPSL-CM4 (Marti et al., 2005) | Institute Pierre Simon Laplace, France | 2.4° × 3.75° |
| MIROC-HIRES (K-1 Model Developers, 2004) | CCSR/NIES/FRCGC, Japan | T106 |
| PCM1 (Meehl et al., 2004) | National Center for Atmospheric Research, USA | T42 |
| HADCM3 (Johns et al., 2003) | Hadley Centre for Climate Prediction and Research/Met Office, UK | 2.75° × 3.75° |
| HADGEM1 (Johns et al., 2006) | Hadley Centre for Climate Prediction and Research/Met Office, UK | 1.25° × 1.875° |

On the other hand, the westerly disturbances reaching the region result from the southern flank of the storm track transporting westward extratropical cyclones. The depressions reaching as far east as the region we discuss here, typically originate or reinforce over the Caspian and the Mediterranean Sea, at the easternmost extremity of the Atlantic and Mediterranean storm tracks (Hodges et al., 2003; Bengtsson et al., 2006). Such western disturbances typically move eastward along the Karakoram and the Himalayan Arc and eventually weaken and die over northern India and south of the Indian subcontinent.

The hydrology of the Ganges, Brahmaputra and Mekong basins is dominated by the summer monsoonal precipitation (Annamalai et al., 2007), with negligible contributions coming from the evanescent winter extratropical weather systems. This fact is evident from the relatively smaller snow and glacier melt runoff generated from the Ganges and Mekong basins (Immerzeel et al., 2010; Hasson et al., 2013a). In contrast, the Indus Basin hydrology is dominated during winter and spring seasons (peak in March) by the influence of western disturbances, mainly in the form of solid precipitation (Hasson et al., 2013b), while in summer (peak in July/August) the monsoonal rainfall contributes critically to the water budget of the basin (Wake, 1987; Rees and Collins, 2006; Ali et al., 2009). Hence, the Indus Basin, located at the boundary between two different large-scale circulation modes, has a more complex hydro-climatology than the other three basins.

## 3 Data and methods

### 3.1 Data sets

In the present study, we have chosen the PCMDI/CMIP3 climate model simulations (Table 1) to investigate the skill of CMIP3 climate models for providing an adequate representation of the hydrological cycle over the considered river basins. Our auditing and verification of the GCMs are based on the historical simulations of 20th century climate (1961–2000) under the present-day climate forcing, whereas the future changes are extracted for the corresponding time spans of the 21st (2061–2100) and 22nd centuries (2161–2200) under the SRESA1B scenario, which is, broadly speaking, a median of all the SRES scenarios. The monthly climatology of the hydrological quantities such as Precipitation ($P$), Evaporation ($E$) and Total Runoff ($R$) are considered for the analysis. The surface upward latent heat fluxes are used to compute the evaporation from all models. For the observational data sets, monthly climatology of the basin-integrated precipitation is computed from the University of East Anglia Climatic Research Unit (CRU) Time Series (TS) high-resolution gridded data version 3.2 (CRU, 2012), while the monthly climatology of the historical river discharges ($D$) from the basins are obtained from the Water and Power Development Authority (WAPDA), Pakistan, Arora and Boer (2001) and Jian et al. (2009).

### 3.2 Methods

In order to estimate accurately the basin-wide monthly climatology of the hydrological quantities from the gridded data sets, the Voronoi–Thiessen tessellation method is used (Okabe et al., 2000). In the case of climate model gridded data sets, inconsistencies between the land–sea masks of GCMs and the extracted basin boundaries are carefully adjusted to avoid any systematic negative biases in the computed water balances, due to the spurious inclusion of portions of coastal grid cells of the GCMs representative of areas with sea surface, where high evaporation, low precipitation, and no runoff are found.

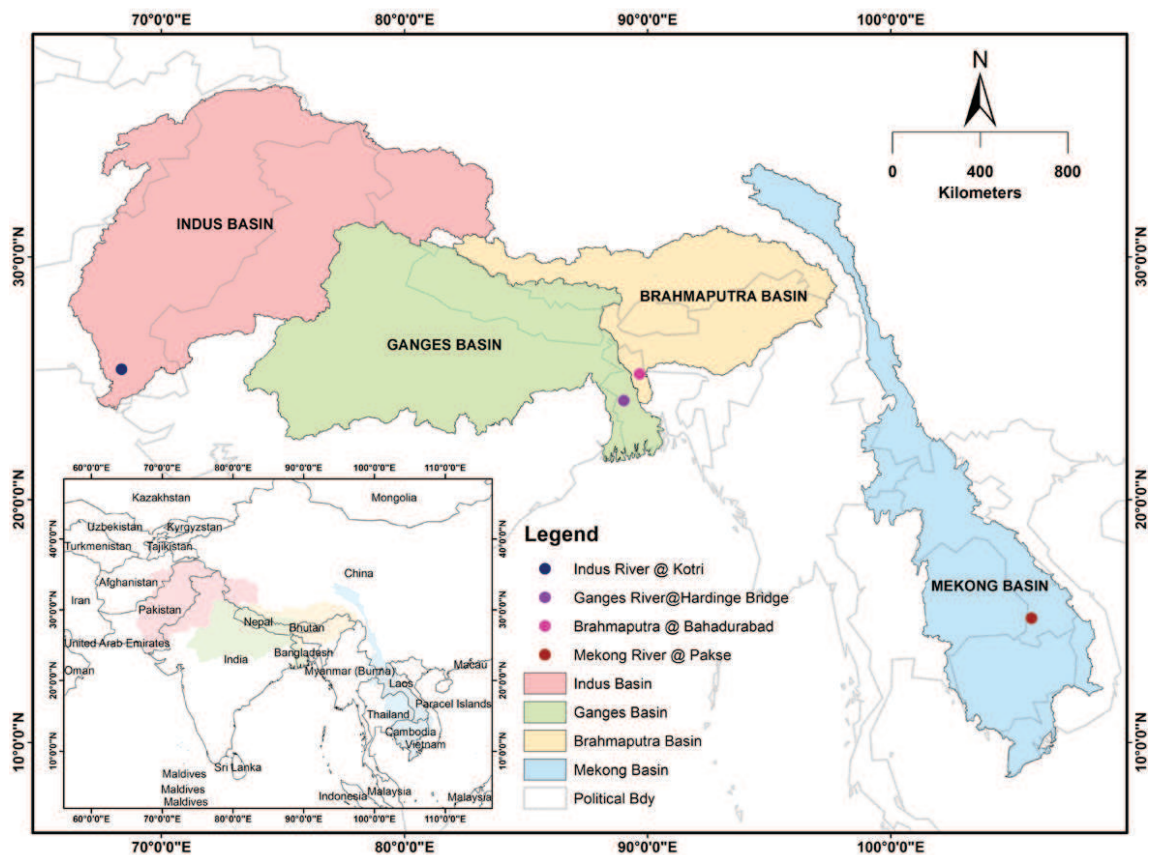

**Fig. 1.** The four river basins considered in this study: Indus, Ganges, Brahmaputra and Mekong (west to east).

Glaciers are currently receding in the Ganges, Brahmaputra and Mekong basins, where, nonetheless, the overall contribution to the runoff of glaciers' meltwater is relatively marginal (Immerzeel et al., 2010; Savoskul and Smakhtin, 2013), The contribution of glaciers' meltwater is, however, extremely relevant in the much drier Indus River basin, which is instead experiencing a relatively stable overall mass balance of its glaciers (Hewitt, 2005; Scherler et al., 2011; Bolch et al., 2012; Gardelle et al., 2012; Bhambri et al., 2013). Therefore, we assume minute corrections of the meltwater runoff, associated with changes in the glaciers' mass, to the historical discharge climatology in all river basins. See our discussion in a previous paper (Hasson et al., 2013a). Following Lucarni et al. (2007, 2008), monthly means of $P$, $E$, $P - E$ and $R$ are computed carefully depending on the adjusted land–sea mask for each GCM, for the considered time spans of the last 40 yr of the 20th, 21st and 22nd centuries using the following equation:

$$\overline{\beta_i} = \frac{1}{A} \int_A \mathrm{d}x \mathrm{d}y \beta_i, \tag{1}$$

where $\beta_i i = 1, \ldots, 4$ corresponds to the mean monthly climatology of each of the four variables mentioned above, $A$ denotes the area of each considered river basin, and $\overline{\beta_i} i = 1, \ldots, 4$ corresponds to the basin-integrated monthly climatology of all considered variables. Our presented approach is equally applicable to the data sets of any resolution or for any other geographical region providing opportunities of its future applications.

In order to characterize the monsoonal precipitation regime for each river basin, we consider its four basic features – onset time, retreat time, duration and magnitude – which are estimated from the models' simulated and observed precipitation data sets. As stated above, the monsoon *onset* is characterized by an abrupt precipitation increase persisting throughout the monsoon season. In order to quantify such an abrupt increase, we estimate a uniform threshold from the normalized observed monthly basin-integrated precipitation, which is calculated as

$$\hat{P}_i = \frac{(P_i - P_{\min})}{(P_{\max} - P_{\min})}, \tag{2}$$

where $i = 1, \ldots, 12$ indicates month of the calendar year, $P_i$ is the basin-integrated total precipitation for the month $i$, $P_{\min}$ and $P_{\max}$ are the minimum and the maximum basin-integrated precipitation during the calendar year, respectively. We found that the onset month, estimated from the observed normalized basin-integrated precipitation, is the $i$th month satisfying the condition $\hat{P}_i - \hat{P}_{i-1} \geq 0.17$. We have verified that this condition gives us the realistic climatic onset months (Janowiak and Xie, 2003; Krishnamurti et al., 2012) for each basin using the observational data. Since the monsoon retreat is fairly smooth, such a condition is not equally applicable for estimating its timings. Therefore, taking the total precipitation of the onset month $i$ ($P_{\text{onset},i}$) as a threshold, we define the monsoon retreat month at the falling limb of the annual cycle of precipitation as a $j$th month such that the condition $P_{j+1} \approx P_{\text{onset},i-1}$ is satisfied. The *duration* of the monsoonal precipitation regime is then defined as the period (in our case number of months) going from the onset month $i$ up to the month of its retreat, $j$. The *magnitude* of the monsoonal precipitation regime is taken as the total amount of precipitation within the monsoon duration. Such a procedure is adopted for the models' simulated precipitation regimes to estimate their suggested timings of the onset and retreat months as well as the monsoon duration and its magnitude for each basin. Of course, using monthly data we derive the properties of the monsoon only with rather coarse time resolution, but the proposed approach is sufficient for the goals of this study.

In order to assess the realism of the models, we have compared their simulated annual cycles of precipitation and runoff with the available observations. Since no CMIP3 climate model implements irrigation, the actual discharges from the highly irrigated river basins cannot directly be used for comparison with the simulated runoffs. This is especially problematic for the Indus Basin for which the observed discharge into sea at the near-to-sea gauging station substantially underestimates the real discharge because of heavy water diversion for irrigation. Such an issue is relatively less important for the other three river basins. In order to reconstruct the real discharge from the Indus Basin, the amount of diverted water and its annual distribution has to be taken into account. Laghari et al. (2012) reports that approximately 170 mm yr$^{-1}$ of water is diverted annually within the Indus Basin. The maximum diversion occurs from the start of the snowmelt season (March/April) till the start of the monsoon season (June/July). This is also evident from the anthropogenically unperturbed discharges at the river inflow measurement stations (RIMs) of the main Indus River at Tarbela and its tributaries (Jhelum at Mangla and Kabul at Nowshera in Pakistan), collected from WAPDA. In view of this fact, the annual amount of diverted water has been redistributed throughout the year according to the total hydrograph of the anthropogenically unperturbed discharges from the mentioned tributaries, and then added to the hydrograph of the observed discharges from the basin.

## 4 Results for 20th century climate (1961–2000)

We first discuss the skill of the GCMs in reproducing the basic properties of the seasonal variability of the hydrological cycle in each considered basin. In Figs. 2–6, the simulated hydrological quantities are shown alongside the observations for the 20th century climate, where each model is coded with a different colour and a marker; the ensemble mean is shown as a dashed black line, whereas the observed climatological quantity is shown as a solid black line. It is worth mentioning here that the common practice to present the arithmetic mean and its standard deviation as the ensemble mean and its spread (Houghton et al., 2001) is intentionally avoided in our analysis. For the reasons described in the introduction, in our analysis we present the so-called ensemble mean values just for indicative purposes.

### 4.1 Indus Basin

*Precipitation:* as discussed in Sect. 2, precipitation over the Indus Basin comes mostly from the South Asian monsoon circulation in summer and from the extratropical cyclones in winter and spring (Hasson et al., 2013b). We therefore expect that the GCMs feature a bimodal precipitation regime for the Indus Basin, showing peaks during the months of March and July. Such a qualitative property is well-represented by most of the models.

For the precipitation regime associated with the western mid-latitude disturbances, there is a fair agreement between models (spread of about 50 mm) in reproducing it, showing correctly that winter and spring are wet seasons for the Indus Basin (Fig. 2a). However, seven out of thirteen models (CNRM-CM3, GISS-AOM, GFDL2, CSIRO3, ECHO-G, PCM1 and HADGEM1) show slightly delayed maxima for the winter/spring season precipitation, suggesting these in April instead of March. The performance of most of the models in reproducing the overall pattern is quite satisfactory, indicating that these models properly simulate the Northern Hemisphere storm track also in this rather peripheral part of it.

Less satisfactory is, instead, the situation for the monsoonal precipitation regime, with most of the models showing serious difficulties in reproducing it. The models remarkably differ with each other in terms of their simulated magnitude and their suggested timings of the onset/departure of the monsoonal precipitation. The most surprising fact is that four models (CGCM2.3.2, GISS-AOM, IPSL-CM4 and INMCM3) have been found unable to capture the monsoon signal at all, showing almost no precipitation during the monsoon season (Fig. 2a). A relatively good performance (skill score $\geq 0.5$) for such models is shown in Sperber et al. (2013). Such skill score (between 0 and 1) is defined based on pattern correlation between the model outputs and the observations. We have considered a skill score of 0.5 and greater (below 0.5) as a high (low) skill while comparing our results with Sperber et al. (2013). This emphasizes the need to revisit the validation/performance of the models within the natural boundaries of the river basins, which are relevant for the water resources management and agricultural impact studies. Further investigations are also needed in order to understand why these models are unable to describe the realistic monsoon precipitation over the Indus Basin as well as in order to find out the factors responsible for such unrealistic features. The CNRM-CM3 model, showing a quite unrealistic pattern of both winter and summer precipitation regimes, does not capture the bimodal precipitation distribution and suggests too strong precipitation in July (twice the observed value, Fig. 2a). However, it exhibits a satisfactory performance in terms of its spatial correlation patterns of monsoonal precipitation climatology and the timings of the monsoon peak and retreat over the whole region (Sperber et al., 2013). We would like to add here that four of the models discussed above (CNRM-CM3, GISS-AOM, IPSL-CM4 and INMCM3) feature serious water balance inconsistencies for the Indus Basin on an annual timescale (Hasson et al., 2013a).

Six out of thirteen models (GFDL2, CNRM-CM3, MIROC-HIRES, HADGEM1, HADCM3 and PCM1) agree on the realistic timings of the monsoon onset, which takes place in the month of July (Fig. 6a). Sperber et al. (2013), analysing the first three of these models, have ranked only GFDL2 high for its skill score in such regards. ECHAM5 shows an early onset, suggesting it in June, whereas two models (CSIRO3 and ECHO-G) delay the onset, suggesting it in August. Only two models (HADGEM1 and PCM1) realistically reproduce the timings of the monsoon precipitation maximum in July, whereas three models (GFDL2, HADCM3 and MIROC-HIRES) feature a one-month delay, thus having it in August (Fig. 2a). These six models, suggesting a realistic monsoon onset, also agree on a smooth retreat in September, consistent with the findings of Sperber et al. (2013). From this, we are led to conclude that models suggesting the realistic timings of the monsoon onset may feature realistic duration of the monsoonal precipitation regime whereas the models suggesting an early (delayed) onset may feature its prolonged (short-lived) duration (Fig. 6b).

*Evaporation:* this quantity is intimately controlled by the soil moisture, insolation, relative humidity of the surface air, and surface winds. Figure 2b shows a better inter-model agreement for evaporation during winter and spring seasons, when minimal evaporation is experienced, than during summer. The disagreement among models, as expected, is much greater during summer. Two models (CGCM2.3.2 and INMCM3) show a negligible evaporation during the monsoon period mainly because these models are unable to simulate the monsoon precipitation regime, so that they represent spuriously dry land. Contrary to this, IPSL-CM4 shows higher evaporation from May to October although it also completely fails to capture the monsoon signals; this may point to some issues in the representation of the land–atmosphere coupling of this model, as already discussed in Hasson et al. (2013a). The GFDL2 model shows the highest evaporation during the monsoon season. Models show quite good agreement regarding the timings of the evaporation maxima in August – a peak flow period – except two models (CNRM-CM3 and PCM1), which simulate the smooth rise of evaporation until July and then its smooth decline until December. From Hasson et al. (2013a), we know that these two models do not conserve water for the Indus Basin, so that a process, such as evaporation, which critically depends on soil water availability, can be seriously affected. The ensemble mean is affected by the very high evaporation given by the CNRM-CM3 model during the monsoon season. As a result, it overestimates the evaporation simulated by most of the models 74 S. Hasson et al.: Seasonality of hydrological cycle in South and southeast Asian River basinsduring such a period, but provides a good approximation for the rest of the year. We remind the reader that we cannot compare the models' outputs to any observational data set.

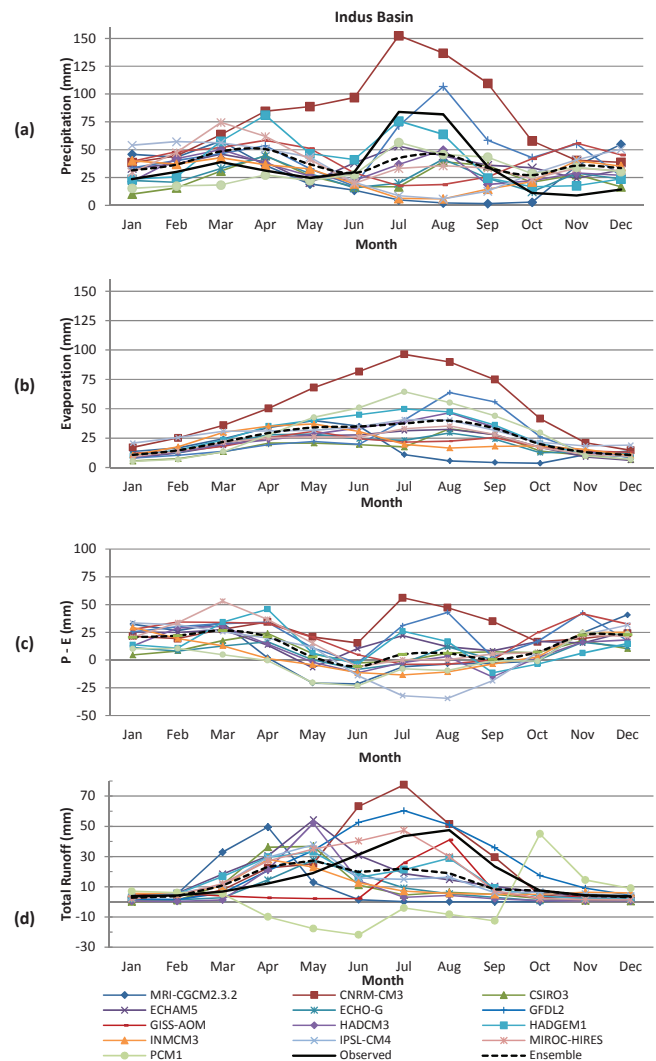

**Fig. 2.** 20th century annual cycles of considered quantities for the Indus Basin: **(a)** $P$, **(b)** $E$, **(c)** $P - E$, **(d)** $R$. Note that as the Indus River is highly diverted for irrigation and other purposes, its estimated natural discharge (by adding diverted volume to actual discharge into sea) is shown.

$P - E$: Fig. 2c shows a negative $P - E$ in June for all models except three (ECHAM5 is characterized by an early monsoon onset, CNRM-CM3 shows overall remarkable precipitation and GISS-AOM delays the realistic spring precipitation maxima). This is associated with the negligible precipitation in the pre-monsoon season and the dependence of the evaporation on the moisture contained in the soil resulting from the snowmelt and the spring precipitation. Even during the monsoon period, only four models (CNRM-CM3, GFDL2, HADGEM1 and ECHAM5) show positive $P - E$ whereas other models suggest it near to zero or negative. IPSL-CM4 model has the lowest value of $P - E$ among models because of the large amount of its suggested evaporation as well as its inability to reproduce the monsoonal precipitation regime. Generally, it is found that the negative or low $P - E$ in the monsoon period is mainly due to the deficiencies in the models in simulating the monsoon precipitation over the Indus Basin. The ensemble mean of $P - E$ suggests almost a null value for the monsoon season whereas it seems to be reasonably representative of the ensemble members during the non-monsoon period. The overall inter-model agreement is better for winter and spring seasons, where a positive water balance is reported, because of the models' better representation of the winter precipitation due to extratropical cyclones and the temperature-constrained low evaporation rate.

*Simulated runoff:* in Fig. 2d we show the annual cycle of the simulated runoff and the observed discharge at the river mouth. In general, these two quantities do not perfectly coincide because the discharge at the river mouth, at any time of the year, results from a non-trivial function of the runoff at previous times, in the various regions of the basin, so that there is a natural time delay between the two quantities. It is a common practice in hydrology to route the runoff using various empirically or physically derived methods (Linear Reservoir approach, Muskingum routing method (Maidment, 1993; Singh and Singh, 2001), Variable Infiltration Capacity – VIC (Liang et al., 1994), TOPKAPI (Konz et al., 2010)) in order to compare it with the observed discharge. In our case, due to the lack of observed runoff data, such routing methods cannot properly be calibrated. However, since we are looking at coarse temporal resolutions (monthly scale) of the basin-integrated total runoff quantity and the typical delay time for such basins is roughly of less than (Ganges, Brahmaputra and Mekong basins) or equal (Indus Basin) to one month, we expect that total runoff ($R$) and discharge ($D$) do not substantially differ. Our assumptions of a typical delay time for the Ganges and Brahmaputra basins are based on the mean travel time reported by Jian et al. (2009). For the Indus and Mekong basins, our assumptions rely on the estimated travel time derived from the flow velocities (Arora and Boer, 1999) and on the knowledge about the basin areas where most of the runoff is generated (Hasson et al., 2013a). Therefore, on the basis of such considerations, we have decided to present the simulated runoff (without applying any routing method) and the observed discharge for each basin in equivalent areal height units (mm).

It is apparent from Fig. 2d that the runoff as simulated by the GCMs is not consistent with the observed discharge of the Indus River. Models show a remarkable spread during the monsoon period, which is associated with their large spread in simulating the monsoonal precipitation. The ensemble mean generally underestimates the observed river discharge. Interestingly, models agree well with each other on the timings of the spring discharge, which is mainly due to the melting of snow accumulated during the winter and in the running spring seasons. This suggests that snow schemes implemented by the CMIP3 climate models perform fairly well. CGCM2.3.2 shows a relatively high discharge in March and April. Surprisingly, we found that PCM1 suggests a negative runoff for the period April to September. Our previous analysis (Hasson et al., 2013a) shows that the water balance of the PCM1 model is fairly closed on an annual timescale ($P - E = R$), with the minimum value among the studied models. However, our present analysis shows that this model simulates a negative runoff from April to September, which further implies serious physical inconsistencies in its water balance. Given the findings of our analysis, the great socioeconomic impact of the Indus water availability and the deficiency found in PCM1 model, we suggest the modellers' communities that the land–surface components of the climate models should be realistically described and tested, particularly for the runoff parameterization schemes.

### 4.2 Ganges

*Precipitation:* the hydrology of the Ganges Basin is dominated by the South Asian summer monsoon, featuring the onset in the month of June, the peak of precipitation from July to August, and the retreat in the month of September (Annamalai et al., 2007). Figure 3a shows the mean monthly precipitation climatology and illustrates the skill of the CMIP3 models in reproducing the Ganges' Basin monsoonal precipitation regime. First, all models except CNRM-CM3 – which shows significant precipitation in early spring – realistically suggest negligible precipitation outside the monsoon period with little differences with respect to the observations. For the monsoon season, the MIROC-HIRES model shows an excellent qualitative as well as a quantitative agreement with the observations. Models realistically reproduce the timings of the onset, retreat, maxima and the overall pattern of the monsoonal precipitation regime. In particular, GFDL2 shows a pattern in very good qualitative agreement with the observations, as also discussed in Sperber et al. (2013). Five models (CGCM2.3.2, ECHAM5, GFDL2, HADCM3 and MIROC-HIRES), which are able to reproduce the timings of the onset (Fig. 6a) exhibit a low skill in Sperber

Earth Syst. Dynam., 5, 67–87, 2014 www.earth-syst-dynam.net/5/67/2014/

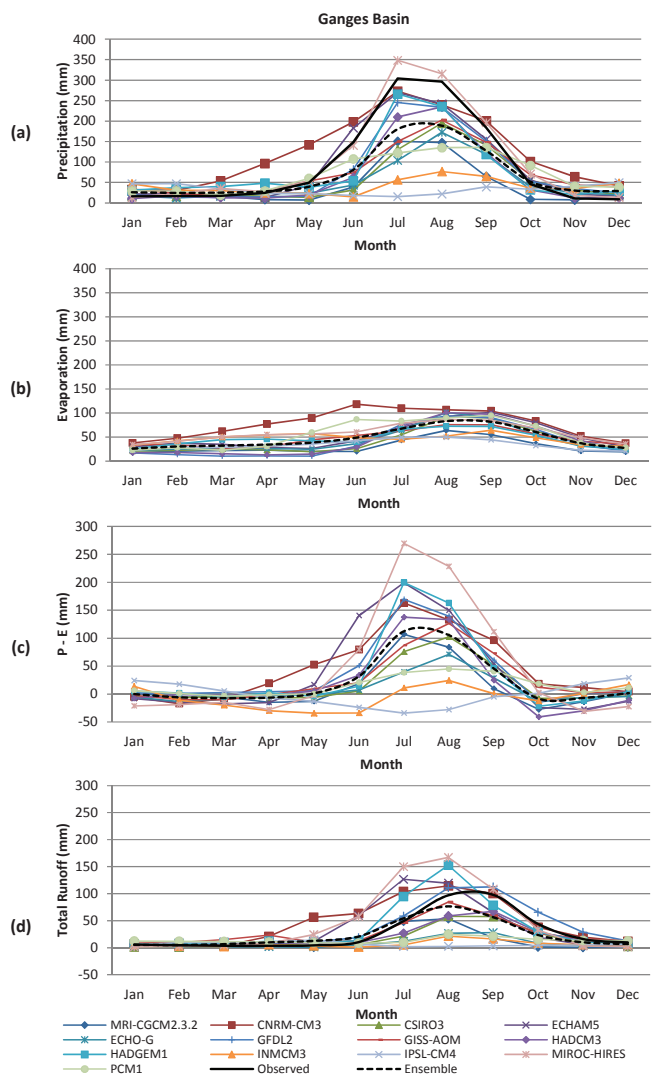

**Fig. 3.** 20th century annual cycles of considered quantities for the Ganges Basin: **(a)** $P$, **(b)** $E$, **(c)** $P - E$, **(d)** $R$.

underestimates the observed precipitation (CRU, 2012) for the Ganges Basin. We have noted that the skill score given in Sperber et al. (2013) does not clearly identify the early and the delayed behaviour of the monsoon onset, peak and retreat for those models which do not realistically simulate such monsoonal characteristics.

*Evaporation*: the seasonal cycle of the evaporation is controlled by the onset and decay of the monsoon, because evaporation is strongly affected by the soil moisture in addition to the insolation and wind speed. All models are in good qualitative agreement in this respect, even if the quantitative agreement is not so good. The CNRM-CM3 model shows the highest evaporation throughout the year, whereas IPSL-CM4 consistently features the lowest (Fig. 3b). The four models (CNRM-CM3, HADGEM1, MIROC-HIRES and INMCM3) show relatively high evaporation in spring, whereas all the models except the five models (GISS-AOM, IPSL-CM4, INMCM3, HADGEM1 and CGCM2.3.2) show relatively low evaporation during the monsoon season. Quite interestingly, the CNRM-CM3 model shows a peak of evaporation (June) before the peak of precipitation because the river basin is quite wet already throughout the spring, so that evaporation in June is not moisture-limited, while the presence of a heavy cloud cover in July and August limits the effect of direct solar radiation at the surface.

$P - E$: good agreement is seen among the models concerning their simulated $P - E$ between November and May, when the precipitation and evaporation compensate up to a good degree of precision. For all models except IPSL-CM4, which, as discussed before, seems to be heavily biased, $P - E$ is positive in the monsoon season, but the values of the excess of precipitation with respect to the evaporation vary wildly (Fig. 3c). Given the large inter-model uncertainties, it seems quite problematic to give any physical interpretation to the ensemble mean.

*Simulated runoff*: the agreement among models is similar to that described above regarding the seasonal cycle of $P - E$. Models agree with each other for the lean flow period as well as for the realistic timings of the peak discharges, which typically occur in August and September with similar magnitudes (Fig. 3d). However, models show a large spread in the magnitude of their simulated runoffs, which can most probably be attributed to the variations in their simulated monsoonal precipitation. CNRM-CM3 suggests a high runoff during April and May as well, as it describes a spurious regime of strong spring precipitation. The simulated runoffs from the five models (MIROC-HIRES, CNRM-CM3, GFDL2, HADGEM1 and ECHAM5) generally tend to overestimate the overall observed discharge, either showing an early rise or a later drop, or both, of their runoff regime. Conversely, the models underestimating the observed discharge regime either feature a delayed rise in the simulated runoff or an early drop in it, or both. The ensemble mean generally seems to underestimate the observed discharge.

et al. (2013) except the GFDL2 model. However, PCM1 suggests an early onset and the prolonged monsoon duration by one month. Five models (CSIRO3, ECHO-G, GISS-AOM, HADGEM1 and INMCM3) suggest a delayed onset, with the duration shortened by one month. Our findings of a delayed onset as suggested by these five models are consistent with Sperber et al. (2013). For the monsoon retreat, all models show it realistically in September. The rest of the models that suggest a realistic timing of the onset also suggest realistic monsoon duration. Surprisingly, IPSL-CM4 model is unable to capture the monsoonal signal at all, just as in the case of the Indus Basin. The CNRM-CM3 model suggests an unrealistically early onset in the month of April. All models suggest that the precipitation peaks either in July or in August, except PCM1, which suggests it in September. Ensemble mean precipitation is placed well in the middle of all models but it

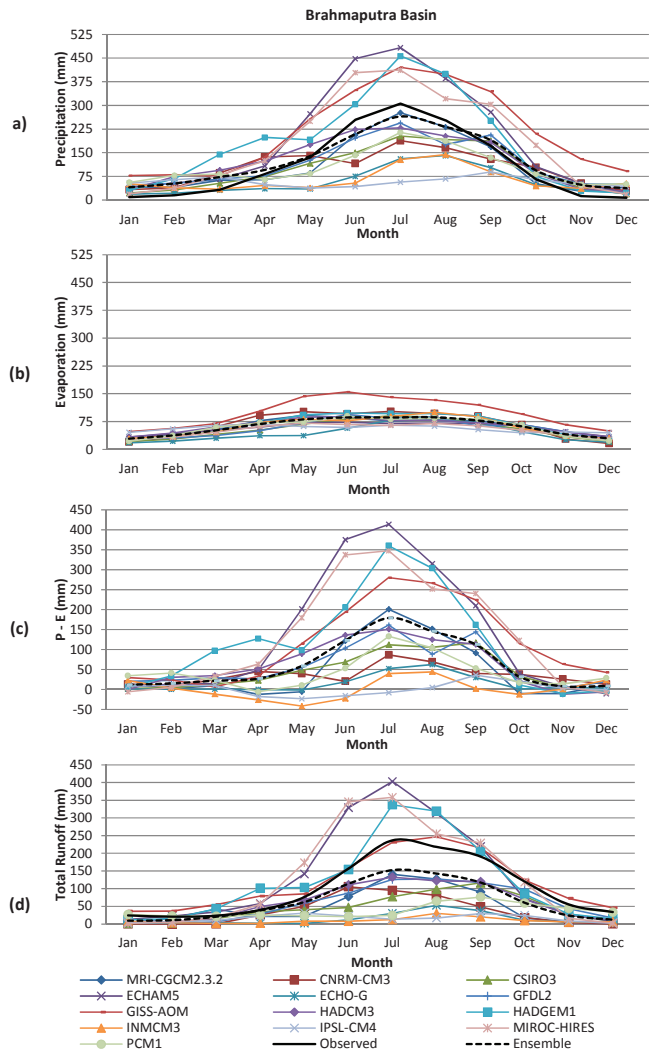
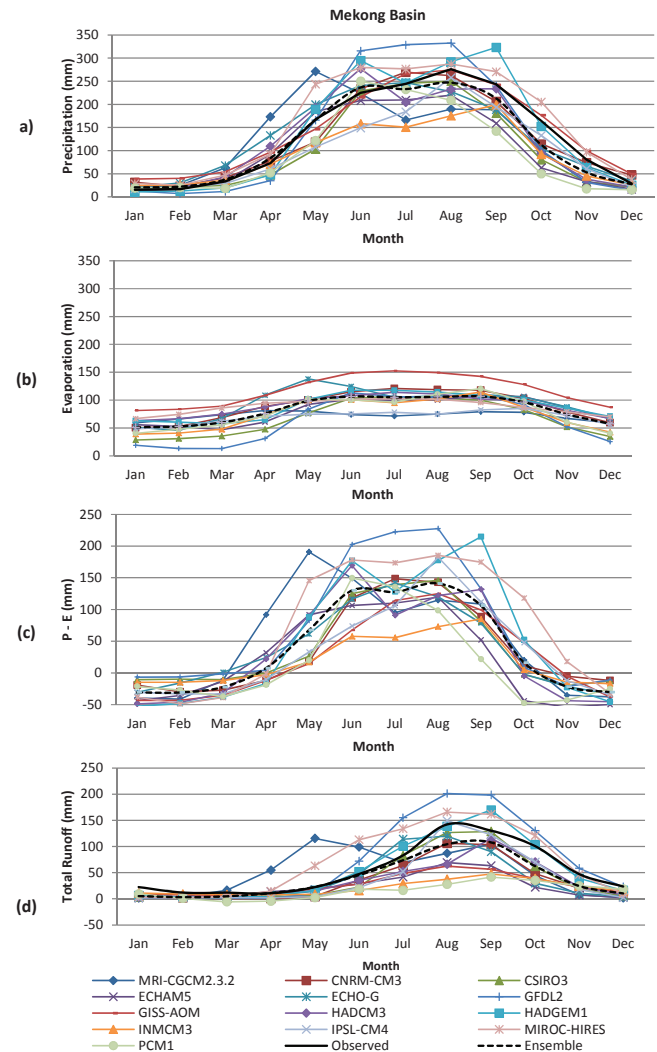

**Fig. 4.** 20th century annual cycles of considered quantities for the Brahmaputra Basin: **(a)** $P$, **(b)** $E$, **(c)** $P - E$, **(d)** $R$.

**Fig. 5.** 20th century annual cycles of considered quantities for the Mekong Basin: **(a)** $P$, **(b)** $E$, **(c)** $P - E$, **(d)** $R$.

### 4.3 Brahmaputra

*Precipitation*: the Brahmaputra Basin's precipitation is mostly associated with the summer monsoon circulation. However, part of its precipitation comes from the small-scale isolated thunderstorms during the pre-monsoon (late spring) season (these processes cannot be directly represented in climate models) and an even smaller quantity comes through winter extratropical cyclones impacting the northern part of the basin. There is a very good inter-model agreement at a qualitative level: all models describe a precipitation regime where the peak is broader than for the Ganges Basin. All models agree at a quantitative level regarding the winter precipitation, while there is a large inter-model variability concerning the magnitude of the monsoonal and pre-monsoonal precipitation (Fig. 4a). IPSL-CM4 again fails to capture the monsoon signal, whereas two models (INMCM3 and ECHO-G) suggest the lowest monsoonal precipitation amongst all CMIP3 models. HADGEM1 shows the highest precipitation in the pre-monsoon season, whereas ECHAM5 suggests the overall highest precipitation regime for the basin. Models seem to agree on the timings of the onset in May. Our analysis suggests that five models (CSIRO3, ECHAM5, MIROC-HIRES, HADCM3 and GFDL2) realistically simulate the timings of the onset and retreat, and so, the monsoon duration (Fig. 6a). For the first three models, such performance is not consistent with Sperber et al. (2013), who show their low skill in capturing the timing of the monsoon onset over the whole region. Four models (CGCM2.3.2, ECHO-G, HADGEM1 and PCM1) delay the onset by one month, whereas two models (INMCM3 and CNRM-CM3) delay it by two months. The findings of Sperber et al. (2013) confirm such behaviour for the four models(CGCM2.3.2, ECHO-G, INMCM3 and CNRM-CM3) though without quantifying their delays.

All models except three (INMCM3, ECHO-G and IPSL-CM4) predict the monsoonal precipitation maxima in July, which is also consistent with what has been found by Sperber et al. (2013). A good agreement is found as far as the retreat is concerned (i.e. September) except for two models (MIROC-HIRES and GISS-AOM), which have a delayed decay in the month of October. Such a feature is not visible in the investigation by Sperber et al. (2013). The GCMs generally underestimate the observed precipitation (CRU, 2012) during the monsoon season but overestimate it during the rest of the year. The ensemble mean is rather different from most of the model outputs, because it is strongly biased by the four very wet models (ECHAM5, MIROC-HIRES, GISS-AOM and HADGEM1). The ensemble mean also underestimates the observed precipitation during the monsoon season but overestimates it during the rest of the year.

*Evaporation:* there is a very good inter-model agreement throughout the year for the simulated evaporation, with the only exception being the GISS-AOM model, which features the highest evaporation (Fig. 4b). Little inter-model variability exists only in the pre-monsoon season. IPSL-CM4 agrees with other models in terms of its simulated evaporation, though it does not capture the monsoon signals for the basin. This confirms the deficiencies in the model in reproducing an accurate water balance, as discussed in the case of the Indus Basin.

$P - E$: except HADGEM1, all models show a good agreement for their computed $P - E$ during the dry season, whereas their large differences exist during the monsoon season, resulting from the large discrepancies found for the precipitation fields (Fig. 4c). Four models (GISS-AOM, ECHAM5, MIROC-HIRES and HADGEM1) suggest relatively high $P - E$ in the monsoon season, whereas IPSL-CM4 shows almost null $P - E$ throughout the year. Very few models suggest slightly negative $P - E$ in the pre-monsoon season. As in the case of precipitation, the ensemble mean also overestimates $P - E$ with respect to most of the models.

*Simulated runoff*: models show a good agreement for the lean flow period (late November to late March) as well as for the start of the high flow period (in the month of April) (Fig. 5d). Two models (INMCM3 and IPSL-CM4) show a negligible runoff throughout the year, whereas two models (ECHAM5 and MIROC-HIRES) are characterized by an early (delayed) rise (drop) in flows. Most of the models substantially underestimate the observed discharge. Interestingly, the observational data set suggests that the four wet models are closer to reality than the others, which are clustered together towards dry conditions. Obviously, the ensemble mean also underestimates the observed discharge throughout the high flow period.

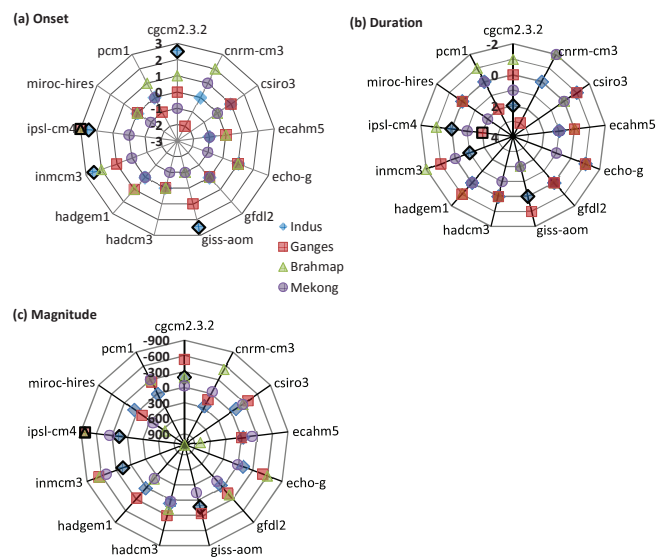

**Fig. 6.** Difference for the 20th century climate (1961–2000) relative to the observations, **(a)** monsoon onset timings, **(b)** monsoon duration, **(c)** monsoonal precipitation magnitude. Positive values indicate the delayed onset and prolonged monsoon duration in months and an overestimation of monsoonal precipitation magnitude in mm, while the negative values indicate the opposite. Note that GCMs which do not simulate the monsoon precipitation regime at all for the particular basin (markers with black border) are only shown for indicative purposes.

### 4.4 Mekong

*Precipitation*: the hydrological regime of the basin is governed by the northeasterly and the southwesterly monsoonal winds. The models show a good agreement concerning a realistic timing of the onset, retreat and duration of the monsoon (Fig. 6a). A lower but still satisfactory inter-model agreement appears with respect to the observed magnitude (usually a slight underestimation is found) of the monsoon precipitation. Our results show that the five models (CSIRO3, GFDL2, INMCM3, IPSL-CM4 and PCM1) suggest realistic timings of the onset in May, however, two of these models (CSIRO3 and INMCM3) feature quite a low skill score for the onset timing of the monsoon in Sperber et al. (2013). Six models (ECHAM5, ECHO-G, GISS-AOM, HADCM3, MIROC-HIRES and CGCM2.3.2) suggest an early onset by one month. These models also feature a low skill score for the onset timing in Sperber et al. (2013). Only CNRM-CM3 shows a delayed onset in June. For the monsoon retreat, models generally suggest an early drop as compared to the observations. Five models (CSIRO3, GFDL2, PCM1, HADGEM1 and INMCM3) suggest the realistic retreat and duration of the monsoon as compared to the observations, whereas five models (ECHAM5, ECHO-G, HADCM3, IPSL-CM4 and CGCM2.3.2) suggest a monsoon duration one month longer than observations, while the overestimate is larger for two models (GISS-AOM and MIROC-HIRES). Only CNRM-CM3 underestimates the monsoon duration by two months. The results from Sperber et al. (2013) indicate that, except ECHO-G, all models show better skill in reproducing the realistic timing of the monsoon retreat rather than the timing of the onset and duration of the monsoon. Moreover, the performance of the models in simulating the realistic monsoon duration is generally associated with their skill in reproducing the right timing of the monsoon onset.

Three models (CGCM2.3.2, HADCM3 and HADGEM1) suggest low precipitation in the month of July, showing two peaks of precipitation, one in the month of June and one in September. Two models (IPSL-CM4 and INMCM3) suggest the lowest monsoon precipitation for the Mekong Basin, thus confirming their dryness throughout the investigated region, whereas GFDL2 features the highest monsoonal precipitation. The ensemble mean, in this case, places itself in the middle of the models' range.

*Evaporation*: this basin appears to be the wettest of the four considered here, because evaporation is relatively high throughout the year and with modest difference between the monsoon season and the rest of the year. Most of the CMIP3 models feature a good degree of agreement in the representation of the evaporation throughout the year, with one model standing out as having the highest evaporation in most months (GISS-AOM). Instead, the outliers with the lowest evaporation are the GFDL2 model in the first five months of the year and the two models (IPSL-CM4 and CGCM2.3.2) for the monsoon season.

$P - E$: the intra-annual variations of $P - E$ (Fig. 5c) is rather similar to that of precipitation because the seasonal cycle of the evaporation is weak. There is a good agreement among all models on the negative $P - E$ in the dry period, i.e. before April and after October. GFDL2 suggests the overall highest amount of $P - E$ whereas INMCM3 suggests the lowest amount. Just as in the case of precipitation and evaporation, ensemble mean places itself in the middle of the models' range.

*Simulated runoff*: as most of the runoff is generated over the lower Mekong Basin and consequently, having a shorter travel distance to the basin outlet/sea, one expects a shorter time delay between the basin-integrated runoff and the discharge of the Mekong Basin. Therefore, our comparison at the face value between the two quantities is indeed meaningful for the Mekong Basin. In Fig. 5d the annual cycle of the monthly mean simulated runoff shows an excellent intermodel agreement during the lean flow period (December to late March). Although there are large differences among few models in terms of their simulated magnitude during the high flow period, most of the models show a good agreement in reproducing the timings of the rise in the runoff in May, the maxima during August/September and the drop in October. Most of the models underestimate the observed discharge regime of the basin, particularly in the early part of the high flow period due to the delayed rise in flows, so the ensemble-mean also underestimates it. Our results show that the overall pattern of the discharge regime for the Mekong Basin is simulated well with some models fairly close to the observations.

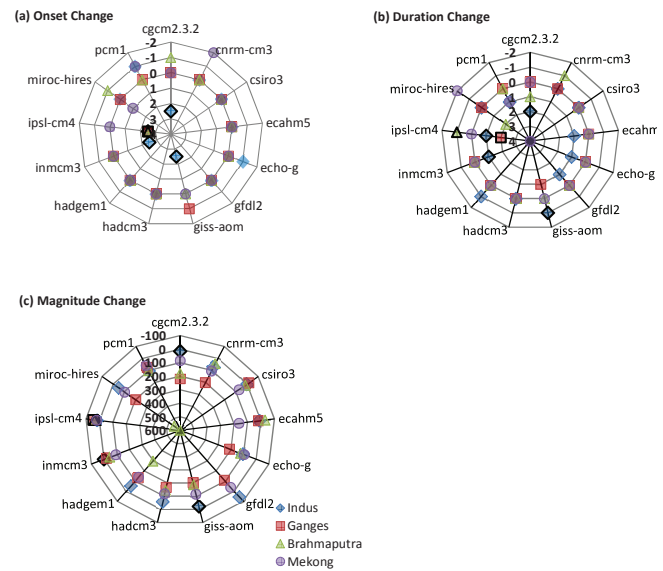

**Fig. 7.** Change for the 21st century climate (2061–2100) relative to 20th century climate (1961–2000), **(a)** in the timings of the monsoon onset month, **(b)** in the monsoon duration, **(c)** in magnitude of the monsoonal precipitation in mm. Positive values indicate the delayed onset and prolonged monsoon duration in months and increased magnitude of monsoonal precipitation in mm, while the negative values indicate the opposite. Note that GCMs which do not simulate the monsoon precipitation regime at all for the particular basin (markers with black border) are only shown for indicative purposes.

## 5 Response to climate change

The overall satisfactory performance of most of the models, in particular for the easternmost basins, suggests studying their simulated changes in the hydrological cycle associated with the increase in the GHG forcing. Hence, we have analysed the impact of climate change in the intra-annual variability of the hydrological quantities discussed above for the 21st (13 models) and 22nd (10 models) century climates relative to the corresponding 20th century climate under the SRESA1B scenario. Since the changes evidenced in the 22nd century climate (2161–2200) are qualitatively very similar (and quantitatively slightly more pronounced) to the changes found for the second half of the 21st century (2061–2100), only the latter are reported in Figs. 7 and 8 and Table 2. Some specific features of the climate change as manifested in the 22nd century are discussed separately. Moreover, we discuss in the text all quantities such as precipitation ($P$), evaporation ($E$), $P - E$ and runoff $R$, but we only show the latter two in Fig. 8, as we have found that $P - E$ illustrates in more significant ways the changes in the hydrological cycle than precipitation and evaporation taken individually. The statistically significant changes in $P - E$, $R$ and in the magnitude of monsoonal precipitation are shown as italic in Table 2.

**Table 2.** Seasonal changes (mm) in $P - E$ and $R$ and changes in the magnitude of monsoonal precipitation for the 21st century climate (2061–2100) as compared to 20th century climate (1961–2000). Note that statistical significance is calculated using Student's $t$ test and statistically significant changes are marked as italic. Negative sign indicates decrease in quantities in mm and vice versa.

|  | Basin→ | Indus | | | | Ganges | | | | Brahmaputra | | | | Mekong | | | |
|---|---|---|---|---|---|---|---|---|---|---|---|---|---|---|---|---|---|
|  | Model↓ | DJF | MAM | JJA | SON | DJF | MAM | JJA | SON | DJF | MAM | JJA | SON | DJF | MAM | JJA | SON |
| $P - E$ | CNRM-CM3 | −14 | −8 | 44 | 4 | −24 | 30 | 81 | 58 | −16 | 17 | 35 | 18 | −17 | 11 | 53 | −7 |
|  | CGCM2.3.2 | 14 | −26 | 7 | 1 | −24 | −10 | 105 | 3 | −41 | 11 | 60 | 49 | −9 | −21 | 31 | 20 |
|  | CSIRO3 | 4 | −10 | −2 | 3 | 0 | −3 | −5 | 0 | −5 | −7 | −21 | −3 | −8 | 0 | 42 | 23 |
|  | ECHAM5 | −27 | −8 | −4 | −2 | −32 | −5 | −58 | 37 | −14 | 14 | −64 | −24 | −12 | *21* | *98* | −2 |
|  | ECHO-G | −11 | −20 | 29 | −3 | −12 | 1 | 77 | 29 | −6 | −2 | 80 | 16 | 2 | 21 | 46 | −4 |
|  | GFDL2 | −10 | −11 | −56 | −6 | −10 | 0 | 108 | 32 | −30 | 25 | *597* | 17 | −4 | −20 | 12 | 51 |
|  | GISS-AOM | 13 | −1 | 14 | 8 | −4 | −4 | 69 | 34 | 15 | 38 | 13 | 71 | −18 | 29 | 33 | −4 |
|  | INMCM3 | −30 | −19 | −7 | −18 | −27 | −7 | 31 | −26 | −44 | −3 | 38 | −24 | −1 | −4 | 20 | −8 |
|  | IPSL-CM4 | 30 | −21 | −14 | −29 | −18 | −23 | 3 | −22 | −23 | −21 | −26 | −31 | −15 | −32 | −20 | 62 |
|  | MIROC-HIRES | −6 | −12 | 9 | 5 | −17 | −29 | *128* | 53 | −19 | *184* | *397* | 71 | −34 | −13 | 76 | 33 |
|  | PCM1 | −5 | −13 | 41 | −12 | −9 | 4 | 47 | 9 | −31 | −8 | *97* | 14 | −9 | 2 | 17 | −1 |
|  | HADCM3 | −7 | −4 | 40 | −7 | −4 | 6 | 94 | 23 | 24 | 26 | *103* | −8 | *15* | 53 | −8 | 31 |
|  | HADGEM1 | −3 | −29 | 42 | −12 | −5 | 2 | 93 | 46 | −2 | *51* | *207* | *60* | *41* | *54* | 17 | *97* |
| $R$ | CNRM-CM3 | 3 | 28 | 37 | 18 | −3 | 36 | 63 | 75 | 4 | 56 | 77 | 56 | −3 | 12 | 16 | 3 |
|  | CGCM2.3.2 | 17 | −25 | 0 | 2 | 3 | −4 | 53 | 20 | 4 | −17 | 49 | 39 | 2 | −17 | 19 | 16 |
|  | CSIRO3 | 1 | −12 | 0 | 3 | 2 | −5 | −1 | −5 | 3 | −17 | −18 | −7 | −3 | −5 | 31 | 27 |
|  | ECHAM5 | 2 | −23 | −25 | 4 | −4 | −13 | −52 | 12 | 0 | −16 | −64 | −14 | 0 | −2 | *75* | 28 |
|  | ECHO-G | 1 | −9 | −8 | 9 | 1 | 2 | 45 | 46 | 1 | 1 | 65 | 22 | −1 | 4 | 50 | 15 |
|  | GFDL2 | 1 | 3 | −55 | −31 | 6 | −3 | 75 | 53 | 18 | 0 | *369* | *217* | 9 | 1 | −24 | 52 |
|  | GISS-AOM | 2 | 1 | *231* | 30 | 27 | −14 | 55 | 44 | 57 | −15 | 56 | 71 | 2 | 4 | 19 | 20 |
|  | INMCM3 | −1 | −37 | −14 | -5 | −7 | −8 | 12 | 3 | −10 | −3 | 9 | 0 | 0 | −2 | 7 | 10 |
|  | IPSL-CM4 | 45 | 1 | *112* | 29 | −2 | −18 | −7 | −6 | 8 | −22 | −16 | −8 | −3 | −1 | −50 | 65 |
|  | MIROC-HIRES | −4 | 7 | −10 | 7 | 3 | −3 | 90 | 47 | 3 | *202* | *325* | *101* | −5 | *37* | −26 | 22 |
|  | PCM1 | −1 | 0 | 10 | 1 | 7 | 0 | 17 | 21 | 9 | −6 | 23 | 48 | −2 | 1 | 4 | 9 |
|  | HADCM3 | 4 | −3 | 10 | *11* | 7 | 0 | 70 | 57 | *11* | 40 | 80 | 30 | 5 | 3 | 50 | 36 |
|  | HADGEM1 | 4 | −15 | 20 | 7 | 3 | −6 | *103* | 36 | 4 | 44 | *214* | *85* | *17* | *13* | 90 | *95* |
|  |  | Magnitude | | | | Magnitude | | | | Magnitude | | | | Magnitude | | | |
| Monsoonal precipitation | CNRM-CM3 | *74* | | | | *202* | | | | 29 | | | | *104* | | | |
|  | CGCM2.3.2 | 15 | | | | *220* | | | | *178* | | | | 86 | | | |
|  | CSIRO3 | 3 | | | | −16 | | | | 15 | | | | 72 | | | |
|  | ECHAM5 | 12 | | | | 19 | | | | −31 | | | | *162* | | | |
|  | ECHO-G | *88* | | | | *210* | | | | *124* | | | | 95 | | | |
|  | GFDL2 | −58 | | | | 109 | | | | *602* | | | | 38 | | | |
|  | GISS-AOM | 22 | | | | *203* | | | | *186* | | | | *115* | | | |
|  | INMCM3 | −1 | | | | 18 | | | | 46 | | | | 92 | | | |
|  | IPSL-CM4 | −16 | | | | −3 | | | | −38 | | | | −20 | | | |
|  | MIROC-HIRES | *49* | | | | *204* | | | | *546* | | | | *126* | | | |
|  | PCM1 | *105* | | | | 80 | | | | *102* | | | | 58 | | | |
|  | HADCM3 | *58* | | | | *168* | | | | *146* | | | | *117* | | | |
|  | HADGEM1 | *42* | | | | *131* | | | | *297* | | | | *138* | | | |

## 5.1 Indus Basin

Almost all models suggest no change in the onset of the monsoon, except ECHO-G. As for the duration of the monsoon, three models (ECHO-G, GFDL2 and ECHAM5) suggest its expansion by one month by the end of the 21st century, which is associated with an early onset suggested by the first model and a delayed retreat suggested by the latter two models. Only HADGEM1 suggests a reduction of the monsoon duration by one month due to an early retreat (Fig. 7). As far as $P$ is concerned, all models agree on a decrease in spring (except CNRM-CM3 and GISS-AOM) and winter precipitation (except GISS-AOM, CGCM2.3.2, CSIRO3 and IPSL-CM4) and on an increase in summer (except GFDL2, ECHAM5, INMCM3 and IPSL-CM4) and autumn precipitation (except GFDL2, HADGEM1, INMCM3 and IPSL-CM4). Large changes are predicted for summer precipitation, with PCM1 suggesting the highest increase and GFDL2 suggesting the highest decrease. For the 22nd century, only HADCM3 shows an opposite sign of change suggesting more precipitation during winter and spring. The overall decrease in the winter and spring season precipitation is consistent with the northward shift of the Atlantic–Mediterranean storm track expected under anthropogenic warming scenarios (Bengtsson et al., 2006) whereas the increase in the fall precipitation seems to be associated with the strengthening of the monsoonal precipitation (Turner and Annamalai, 2012). Furthermore, a decrease in the precipitation regime of the mid-latitude disturbances suggests a reduction in the snowfall, which implies a lower snowmelt contribution to the spring runoff. This is a serious issue since it poses a great risk for the mass balance of existing Hindu Kush–Karakoram–Himalayan (HKH) glaciers, as well as the timings of the water availability downstream (Hasson et al., 2013b).

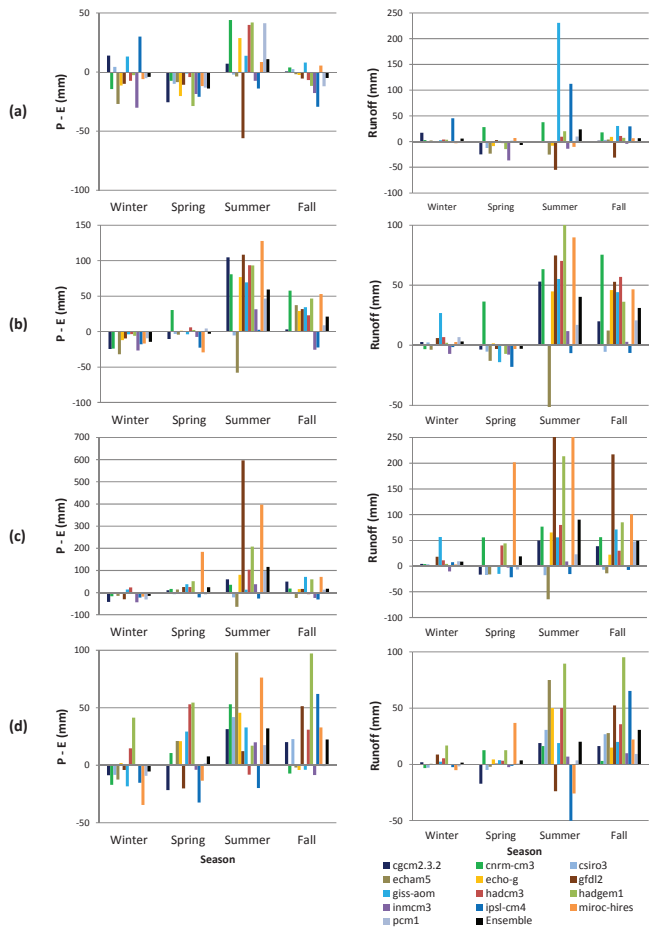

**Fig. 8.** Intra-annual changes in $P - E$ (left) and $R$ (right) for the 21st Century Climate (2061–2100) with respect to 20th Century Climate (1961–2000) for the basins, **(a)** Indus, **(b)** Ganges, **(c)** Brahmaputra, **(d)** Mekong.

It has been found that models largely agree on an increase in evaporation throughout the year. Only three models show contrary behaviour in spring (ECHAM5, ECHO-G, INMCM3), two models in summer (ECHAM5 and GFDL2) and one (GFDL2) in fall season, due to the decreased soil moisture. Concerning $P - E$, there is a good inter-model agreement on the negative change during spring season for both 21st and 22nd centuries except HADCM3 for the 22nd century. Most of the models also agree on the negative change in $P - E$ for winter and fall seasons. For winter, exceptions are four models (CGCM2.3.2, CSIRO3, GISS-AOM and IPSL-CM4) by the 21st century (Fig. 8a). For the fall season, exceptions are three models (CSIRO3, GISS-AOM and MIROC-HIRES) by the 21st century and two models (CSIRO3 and ECHAM5) by the 22nd century. Most of the models suggest a decrease (increase) in spring (summer) $P - E$, which is similar to the ones in the runoffs.

As the basin receives most of its winter and spring precipitation in the form of snow, a decrease in the spring runoff and in spring precipitation implies a reduced snowmelt contribution to the spring runoff, which can further be linked to a decrease in winter precipitation. On the other hand, an increase in winter runoff, but decrease in winter precipitation and spring runoff, also suggests that solid precipitation will partially be transformed into a liquid precipitation in such season. Such precipitation instantly contributes to the winter runoff, reducing the delayed effects of snowmelt to the spring runoff.

### 5.2 Ganges Basin

There is no change in the onset timings nor for the duration of the monsoon among the models, except for GISS-AOM, which suggests an expansion of the monsoon season by one month due to its delayed withdrawal (Fig. 7). Furthermore, there is a good agreement among models concerning the winter precipitation decrease. The only exceptions are three models – CSIRO3, GISS-AOM and HADCM3 – showing contrasting behaviour (Fig. 8). A general increase in summer precipitation is predicted by most of the CMIP3 models, except three models (CSIRO3, ECHAM5 and IPSL-CM4); the same applies for the fall season, except for two models (INMCM3 and IPSL-CM4). No inter-model agreement is evident for the changes in spring precipitation. It is important to remember that INMCM3 and IPSL-CM4 feature very serious water conservation problems for the Ganges Basin on mean annual timescale (Hasson et al., 2013a). The negative change during the dry season and the positive change during the wet season may suggest an intensification of the precipitation regime associated with the summer monsoon.

As seen for the Indus Basin, there is also a consistent behavior among most of the models on a general increase in evaporation in all seasons, which is mainly associated with a rise in temperatures throughout the year under the warmer climates. However, two models (HADGEM1 and IPSL-CM4) in winter, five models (CSIRO3, ECHAM5, INMCM3, HADGEM1 and IPSL-CM4) in spring, four models (CSIRO3, GFDL2, HADGEM1 and IPSL-CM4) in summer and three models (GFDL2, HADGEM1 and IPSL-CM4) in fall show opposite signs, suggesting a decrease in evaporation by the end of the 21st century. For the 22nd century, a small decrease is suggested by four models (GFDL2, HADGEM1, INMCM3 and IPSL-CM4) in winter, three models (CSIRO3, ECHAM5 and INMCM3) in spring, three models (CISRO3.0, GFDL2, HADGEM1) in summer and four models (GFDL2, HADGEM1, INMCM3 and IPSL-CM4) in fall season.

Concerning $P - E$, the models clearly foresee a decrease in winter and spring seasons by the 21st and 22nd centuries, however three models (CNRM-CM3, HADCM3 and HADGEM1) are an exception for spring season of 21st and 22nd centuries and IPSL-CM4 for winter season of the 22nd century. The models also agree on its increase in summer (except CSIRO3 and ECHAM5) and fall (except INMCM3 and IPSL-CM4) by the 21st and 22nd centuries. Sign of change for the runoff quantity is similar to that of $P - E$, except during the winter season, where $P - E$ is negative but models agree well for the increasing tendency of the winter runoff. We speculate that such a scenario is mainly associated with the increased melting during winter season. There is a negative change in spring runoff except in case of CNRM-CM3, which suggests a positive change.

### 5.3 Brahmaputra Basin

Most of the models suggest no major changes in the timing of the monsoon in the 21st century, except the MIROC-HIRES, which suggests an early onset and a delayed withdrawal by one month, the CGCM2.3.2, which indicates an early onset by one month, and the CNRM-CM3, which instead suggests a delayed onset by one month. Almost all models agree on the strengthening of spring, summer and fall precipitations, except two models (ECHAM5 and IPSL-CM4), which indicate such change for the summer and fall seasons only, while the IPSL-CM4 becomes wetter in spring season. There is also a good agreement between models regarding either negligible or small negative change for the winter precipitation except two models (GISS-AOM and HADCM3) for the 21st and only HADCM3 for the 22nd century. The observed decrease in the winter season precipitation for both centuries clearly suggests a reduced role of extratropical cyclones, which in the current climate roughly contributes about 10 % to the total Brahmaputra precipitation as estimated from the observational data set used in the present study. Also, increases in spring and summer precipitation suggest an intensification of the pre-monsoonal as well as the monsoonal precipitation regime of the basin.

Concerning evaporation, most of the models predict an increase throughout the year, showing an even better agreement on such change as compared to other basins for both 21st and 22nd centuries. For the 21st century, ECHAM5 suggests a small decrease for winter and spring evaporation while IPSL-CM4 suggests its decrease in winter, spring and fall seasons. HADGEM1 shows slight negative changes for winter and fall seasons, whereas GFDL2 shows a negative change only for summer season. For the 22nd century, negative change in the evaporation is suggested by IPSL-CM4 for fall and winter seasons, by three models (ECHAM5, INMCM3 and IPSL-CM4) for spring season and by GFDL2 for summer season. Most of the models agree on a negative change in $P - E$ in winter and on a positive change in it for the rest of seasons. Exceptions are two models (GISS-AOM and HADCM3) for winter, three models (CSIRO3, ECHAM5 and IPSL-CM4) for summer, three models (ECHAM5, INMCM3 and IPSL-CM4) for fall and only IPSL-CM4 for spring season for the 21st century. Two models (GFDL2 and MIROC-HIRES) suggest a higher positive change for the summer as compared to other models. We remind the reader here that two of these models (INMCM3 and IPSL-CM4) feature water balance inconsistencies on annual timescales (Hasson et al., 2013a).

Concerning changes in the runoff, there is a good agreement between most of models on an increase in winter runoff despite a decrease in $P - E$, which calls for an increase in snowmelt, while there is no uniform response regarding the spring runoff. The models also agree on a robust increase in the summer runoff, due to the strong increase in $P - E$.

### 5.4 Mekong Basin

Almost all models foresee no considerable changes in the timing of the monsoon, except MIROC-HIRES, which suggests a phase advance as well as the shrinkage of the monsoonal precipitation regime by one month (by simulating an early onset by one month and an early withdrawal by two months), while the CNRM-CM3 model is an outlier, as it foresees a large variation in the local climate, suggesting a two-month delayed onset and a four-month delayed withdrawal.

As for the Brahmaputra Basin, precipitation is consistently shown to increase during spring, summer and fall seasons by the 21st century. Exceptions to such a general trend are six models (CGCM2.3.2, CSIRO3, GFDL2, INMCM3, IPSL-CM4 and MIROC-HIRES) for the spring season, two models (GFDL2 and IPSL-CM4) for the summer season and INMCM3 for the fall season. Most of the models exhibit a decrease in winter precipitation except three (HADCM3, HADGEM1 and INMCM3). It is also worth noting that ECHO-G, which overall shows a good skill in simulating the monsoon precipitation regime, suggests a large precipitation drop for the spring, summer and fall seasons for the 22nd century, in contrast to the 21st century. It implies that the long-term response of the ECHO-G model will be quite different from its short-term response. The evaporation is expected to increase (decrease) in the summer and fall (spring) seasons, whereas no inter-model agreement is found for the 21st century winter seasons. For the 22nd century, a negative (positive) change in winter and spring (summer and fall) seasons is predicted by most of the models. Again, ECHO-G shows a strong negative change in the evaporation during all seasons.

Looking at $P -, E$, the models agree on the positive change in all seasons except for the winter season.

Exceptions are two models (HADCM3 and HADGEM1) for winter, five models (CGCM2.3.2, GFDL2, INMCM3, IPSL-CM4 and MIROC-HIRES) for spring, two models (HADCM3 and IPSL-CM4) for summer and two models (CNRM-CM3 and INMCM3) for fall seasons, showing an opposite sign of change for the 21st century. For the 22nd century, three models (ECHO-G, HADCM3, HADGEM1) in winter, four models (IPSL-CM4, INMCM3, GFDL2, ECHO-G) in spring, two models (IPSL-CM4 and ECHO-G) in summer, and two models (INMCM3 and ECHAM5) in fall suggest an opposite sign of change. In good agreement with that, we find an increase in summer and fall runoffs but no considerable change in the winter and spring runoffs by the end of the 21st and 22nd centuries. Our analysis reveals that the 22nd century response for this basin can be different from what is projected for the 21st century, as opposed to the other studied basin.

## 6 Discussions and conclusions

In this study, we have analysed the skill of CMIP3 coupled climate models in simulating the intra-annual variations of the hydrological cycle of four major South and Southeast Asian river basins (Indus, Ganges, Brahmaputra, Mekong) for the 20th century under present-day climate forcing and for the 21st and 22nd centuries under SRESA1B scenario. As opposed to previous studies based on grid points or country-based averaging (Annamalai et al., 2007; Turner and Annamalai, 2012; Sperber et al., 2013), the focus in this paper has been on the basin scale, which is the natural context for studying the river hydrology. We have verified whether GCMs' simulated precipitation and total runoff are in agreement with the historical observations and whether the models are able to reproduce the basic properties of the monsoonal precipitation and runoff regimes. Our results show that model performances are different for each river basin, depending on their ability to represent the different circulation modes that govern their hydrology.

In particular, the complex interplay of the seasonal precipitation regimes over the Indus Basin is distinctly represented in the models and reflected in their seasonal precipitation distribution. Our analysis shows that most of the models agree reasonably well in reproducing the pattern of the winter/spring precipitation regime associated with the mid-latitude cyclones, whereas their performance in simulating the basic characteristics of the monsoonal precipitation regime (magnitude of the precipitation, timings of its onset, retreat and the maxima) is more limited, with almost no inter-model agreement. Such a result points out the need for more research for this river basin. A better agreement is found in the representation of the hydrology of the Brahmaputra and Mekong basins, as compared to the other basins. It is worth mentioning here that among the GCMs – GFDL2, CGCM2.3.2 and ECHAM5 – showing the realistic monsoon climatology on a regional scale (Annamalai et al., 2007; Turner and Annamalai, 2012) and studied here, only GFDL2 shows a satisfactory performance for all the study basins. Furthermore, only the GFDL2 model predicts the realistic onset timings of the monsoonal precipitation regime for all the four basins. Sperber et al. (2013), focusing on the model performance at a regional scale, also show a good skill of the GFDL2 model for reproducing the timing of the monsoon onset. There appears to be a good inter-model agreement for realistically simulating the runoff during the lean flow period, while there is a modest agreement for reproducing the overall pattern of the flow regimes for all study basins. On the other hand, we have found no inter-model agreement for the simulated runoffs during the monsoon season, which is associated with the inability of the models to accurately simulate the monsoonal precipitation over the region.

It is noted that four models (IPSL-CM4, INMCM3, CGCM2.3.2 and GISS-AOM) have shown no skill at all in capturing the monsoon signals for the Indus Basin and the same applies to one model (IPSL-CM4) for the Ganges and Brahmaputra basins. This is a serious issue that would indeed require further investigations to point out the governing factors responsible for such inaccuracies. Contrary to this, Sperber et al. (2013), while assessing the CMIP3 model performance over the whole South and Southeast Asian summer monsoon domain, show a relatively better performance of these models. A following study in which the basin scale approach will be extended to CMIP5 models should clarify this aspect. Moreover, we have found that the models that reproduce well the monsoonal characteristics here show low skill in Sperber et al. (2013). For example, one model (ECHAM5) for the Ganges Basin, three models (ECHAM5, CSIRO3 and MIROC-HIRES) for the Brahmaputra Basin and two models (CSIRO3 and INMCM3) for the Mekong Basin feature a low skill in reproducing the timings of the monsoon onset in Sperber et al. (2013). Similarly for the monsoon retreat, two models (MIROC-HIRES and GISS-AOM) for the Brahmaputra Basin and one model (ECHO-G) for the Mekong Basin exhibit low skill in Sperber et al. (2013), in contrast to the findings of the present study.

In addition to assessing the regional scale model performances, we stress the need to investigate the model performance over the natural boundaries of the river basins, which are quite relevant for the water resources management, agricultural practices and various impact studies. Moreover, in view of the large spread among the models in their simulated monsoon precipitation regimes, with models that tend to be drier than the observations, we are of the idea that model performances should be investigated by taking into account the relative thresholds – as done in this study – for the identification of monsoon onset and retreat rather than in terms of prescribed rates (e.g. $5\,\mathrm{mm\,day^{-1}}$ used in Sperber et al., 2013). Besides these differences, our results also show consistencies with the results of Sperber et al. (2013) – with four models (CSIRO3, ECHO-G, GISS-AOM, and INMCM3) for the Ganges Basin, four models (CGCM2.3.2, ECHO-G, IN-MCM3 and CNRM-CM3) for the Brahmaputra Basin and five models (ECHAM5, ECHO-G, GISS-AOM, MIROC-HIRES and CGCM2.3.2) for the Mekong Basin showing similar results in terms of delayed timings of the monsoon onset.

Among these, two models (IPSL-CM4 and INMCM3) also see serious water balance inconsistencies for such basins on annual timescale as shown in a companion paper (Hasson et al., 2013a). One surprising fact noted here is that PCM1 erroneously suggests a negative simulated runoff between April and September for the Indus Basin and between March and April for the Mekong Basin. This kind of error could not be discovered in a previous paper dealing with the annual averages of the hydro-climatological quantities in the region (Hasson et al., 2013a) because such inconsistent behaviour was masked in the overall annual budget. In view of this erroneously simulated quantity $R$, we considered it necessary that the land-surface components of the climate models should be realistically described and tested, particularly the runoff parameterization schemes and other relevant quantities that have great societal importance. Moreover, in order to properly validate the model generated runoff and its various components, the observed river discharges at the various places in the basin should more easily be available to the hydro-climatic community.

Despite a fairly good representation of specific precipitation regimes (winter/spring precipitation over the Indus Basin, summer monsoonal precipitation over the Brahmaputra and Mekong basins), CMIP3 models face problems in representing correctly complex regimes with a higher temporal variability and qualitative shifts between different circulation patterns leading to rainfall (interplay of western disturbances and monsoonal rains over the Indus Basin, of convective processes and tropical disturbances over the Ganges and Brahmaputra Basin), largely controlled by orographically altered (sub-scale) atmosphere–surface energy fluxes over the Tibetan Plateau and related variations of the mid- to upper tropospheric circulation modes (Böhner, 2006). In particular, the Indus Basin, placed at the end of both the storm track and the monsoon influenced areas, poses a great challenge to climate models in terms of adequately simulating its complex hydro-climatology.

A second goal reached in this study has been to assess the projections of the CMIP3 models for the 21st and 22nd centuries under SRESA1B scenario. Our results show that for the Indus Basin, models qualitatively suggest an increase in summer and fall precipitation, but it is not easy to assess the reliability of this result standing the above-mentioned problems in describing the monsoonal circulation in this area. However, most importantly, models suggest a decrease in winter and spring precipitation due to the northward shift of the Atlantic–Mediterranean storm track under warmer climate conditions. Therefore, such a robust pattern of climate change not only threatens the renewal of the existing HKH glaciers but will also cause these glaciers to recede due to both the reduced amount of snow and increasing temperatures. This also implies that there would be a smaller snowmelt contribution to the spring runoff under the warmer climates. Given that at present meltwater from the glaciers and snowmelt contributes 60 % to the total discharge of the Indus, assuring a comparatively stable runoff regime, these findings indicate a shift from a rather glacial and nival to a more pluvial runoff regime for the Indus. Although the consequences of the glacier retreat for the runoff regimes of the major Asian river basins have long been overestimated and only recently been reassessed through more reliable estimates of (snow and glacier) meltwater contribution (e.g. Immerzeel et al., 2010), climate change and consequent changes in runoff regimes and related extremes can raise severe implications for the timely water availability for households, industry and irrigated agriculture, especially in the downstream areas of the 'lifeline' Indus (Hasson et al., 2013b). Our analysis clearly suggests that the Indus Basin will become drier (reduced $P - E$) under the warmer climatic conditions.

For the Ganges Basin, models indicate a decrease (increase) in winter (summer and fall) precipitation for the 21st and 22nd centuries, with a weak inter-model agreement during spring season. For Brahmaputra and Mekong basins, models suggest a positive projected change in precipitation during summer, fall and spring seasons however there is a minute negative change during the winter season. This shift in precipitation regime will clearly affect the overall snow accumulation and then the melt contribution to the spring runoff, which is not clear from the ascertained projected changes in it. Overall, the models suggest an increased water availability for the monsoon dominated river basins (Ganges, Brahmaputra and Mekong) under the warmer climate, which is mainly associated with the strengthening of the South and Southeast Asian monsoonal precipitation regimes. The reduced water availability is, instead, foreseen for the Indus Basin which is mainly associated with a decrease in the precipitational regime of the mid-latitude western disturbances due to the poleward shift of the North Atlantic storm track (Bengtsson et al., 2006).

Considering our present investigations based on monthly climatology, we suggest that further analysis should focus on a smaller timescale to study the variability of the onset timings and phase shifts of the annual cycle of the runoff regimes under warmer climate conditions (Lal et al., 2001; Arora and Boer, 2001), which have a significant practical value for the water management as well as for the agricultural applications. Furthermore, spatially differentiated information within each basin would also be of great utility in order to support the management of water resources and the planning of mitigation measures in the areas affected by natural hazards and substantial climate change. Hasson et al. (2013a), together with the present study, provide information on area-integrated hydrological quantities for the catchments with large transboundary extents but cannot provide information on, for example, future upstream-downstream water availability, changing runoff dynamics and related risks (droughts, floods, etc.). Given the practical needs of an integrated watershed management, we suggest supplementing the model simulations by means of dynamical downscaling, nesting a high-resolution regional climate model (RCM) within a coarser-resolution GCM. The state-of-the-art RCMs enable a stepwise (multi-nesting-level) downscaling of GCM outputs down to few kilometers' horizontal grid-mesh resolution, required to account for the topographic heterogeneity in the high mountain environments (Langkamp and Böhner, 2010; Böhner and Langkamp, 2010). We are currently trying to implement a working version of the WRF model (Michalakes et al., 2001) with a domain centred over South Asia. This approach can be further extended by coupling suitable spatially distributed hydrological models such as variable infiltration capacity – VIC (Liang et al., 1994), TOPKAPI (Konz et al., 2010) and semi-distributed hydrological model such as University of British Columbia Watershed Model – UBC WM (Quick and Pipe, 1976; Singh, 1995) and Snowmelt Runoff Model – SRM (Immerzeel et al., 2009) to RCMs/GCMs, in order to improve the assessment of the changes in the hydrology of the river basins as a result of changed climate conditions. Such modelling studies on the impacts of changes of the hydrological cycle in South and Southeast Asia, from basin to local scales, will be the focus of our future work. The results presented here elucidate that, nonetheless, such downscaling procedures make sense only if the large-scale water budget is well-represented by the GCM within which the nesting is performed. Therefore, it seems relevant for the local downscaling communities, currently nesting RCMs with the CMIP3 GCMs, to take into account our results when attempting to construct high-resolution climate scenarios for the region.

Finally, let us mention that in general, climate change is only one of the drivers causing significant changes in the hydrology, as economic and social changes pose multiple pressures on the water resources. This is particularly distinct in Asia, where tremendous land use, structural, and socioeconomic changes are taking place. Such non-climatic factors undoubtedly worsen the severity of the calamities of climate change, as in the case of 2010 flood in Pakistan. The concurrence of these effects stresses the need for integrating climate and land use scenarios when analysing and assessing the future water availability and its variability.

*Acknowledgements.* The authors acknowledge the modelling groups, the Program for Climate Model Diagnosis and Intercomparison (PCMDI) and the WCRP's Working Group on Coupled Modelling (WGCM) for their roles in making available the WCRP CMIP3 multi-model data set. Support of this data set is provided by the Office of Science, US Department of Energy. SH and JB acknowledge the support of BMBF, Germany's Bundle Project CLASH/Climate variability and landscape dynamics in southeast Tibet and the eastern Himalaya during the Late Holocene reconstructed from tree rings, soils and climate modelling. VL and SP acknowledge the support of the FP7/ERC Starting Investigator grant NAMASTE/Thermodynamics of the Climate System (Grant No. 257106). The authors also acknowledge the support from CliSAP/Cluster of excellence in the Integrated Climate System Analysis and Prediction.

Edited by: A. Kleidon